\documentclass[twocolumn,aps,prb]{revtex4-2}

\usepackage{amsfonts}
\usepackage{amsmath}
\usepackage{amssymb}
\usepackage[pdftex]{graphicx}
\usepackage{dcolumn}
\usepackage{bm}
\usepackage{braket}
\usepackage{color}
\usepackage{here}
\usepackage{mathrsfs}
\usepackage{mathtools}
\usepackage[normalem]{ulem}

\begin{document}

\title{Heat flow from a measurement apparatus monitoring a dissipative qubit}

\author{Tsuyoshi Yamamoto}
\email{yamamoto.tsuyoshi.ts@u.tsukuba.ac.jp}
\author{Yasuhiro Tokura}
\affiliation{Faculty of Pure and Applied Sciences, University of Tsukuba, Tsukuba, Ibaraki 305-8571, Japan}

\date{\today}

\begin{abstract}
We investigate the heat flow of a qubit coupled to heat baths under continuous quantum measurement. In the steady-state limit, we show that heat always flows from the measurement apparatus into the qubit regardless of the measured qubit state and derive lower and upper bounds for the heat current between the qubit and the measurement apparatus. Furthermore, we study the transient dynamics of the heat current and the excess heat during the transient regime.
\end{abstract}

\pacs{Valid PACS appear here}

\maketitle

\section{introduction}

Quantum measurement is a fundamental operation in quantum information processing and plays a pivotal role in the operation of quantum computers.
In contrast to classical measurements, quantum measurements alter the state of the measured system, a phenomenon referred to as the backaction of quantum measurements~\cite{Wiseman2009_text, Jacobs2014_text}.
It has spurred extensive research into phenomena unique to the backaction, such as measurement-induced phase transitions~\cite{Li2018, Li2019, Skinner2019, Ippoliti2021, Minato2022} and non-Hermitian dynamics~\cite{Lee2014, Ashida2016, Ashida2018, Hasegawa2022}.

Besides changing the state of the measured system, quantum measurements can generate heat~\cite{Erez2008, Brandner2015, Elouard2017npj, Elouard2017prl}.
The unification between quantum measurements and thermodynamics has led to the development of the field of quantum thermodynamics~\cite{Pekola2015, Benenti2017}.
From this aspect, measurement-based thermal machines, e.g., Maxwell’s demon~\cite{Koski2014prl, Murayama2009} and the Szilard engine~\cite{Koski2014pnas, Pekola2021epn}, have been studied theoretically~\cite{Elouard2017npj, Elouard2017prl, Manikandan2022, Buffoni2019} as well as experimentally in well-controllable setups, using, e.g., superconducting circuits~\cite{Maillet2019, Ono2020, Naghiloo2020} and cold atoms~\cite{Murch2008, Syassen2008, Jayaseelan2021}.

Among these endeavors, heat exchange between the measured system and the quantum measurement apparatus itself is intrinsically driven by the effects of quantum measurements.
The question of how much heat can be extracted or stored in a quantum system due to quantum measurements merits considerable attention from the energetic aspect of quantum information technology.
To exchange heat between the quantum measurement apparatus and the measured systems, feedback control or contact with the environment is necessary, and it has been studied for projective measurement~\cite{Solfanelli2019, Koshihara2023}.
Recently, novel thermal machines have been proposed based on not only projective measurement but also continuous measurement~\cite{Talkner2017, Yanik2022, Bhandari2022, Yamamoto2022, Bhandari2023}.
However, fundamental questions remain unexplored regarding the direction of heat flow between the measured system and the monitoring apparatus itself, i.e., whether the measured system is heated or cooled when the quantum system is coupled to the environment.
This issue holds significance not only from the viewpoint of information thermodynamics but also for managing the challenging heating problem that may arise in practical implementations of quantum computing.

\begin{figure}[tb]
    \centering
    \includegraphics[width=\columnwidth]{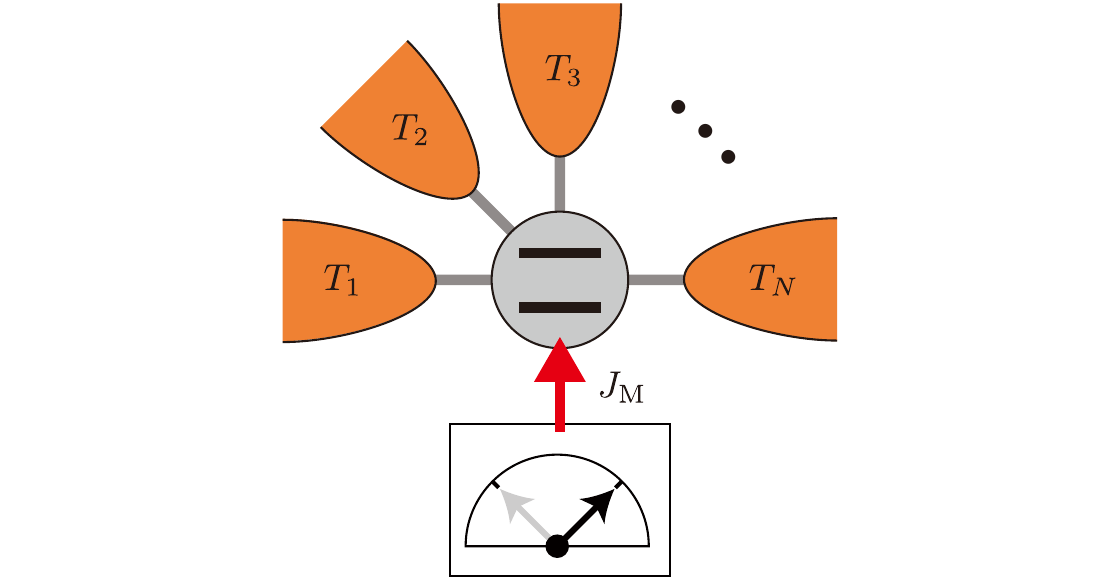}
    \caption{Setup of continuous quantum measurement of a qubit coupled to heat baths with temperatures $T_r$ ($r=1,2,\dots,N$). Continuous quantum measurement of the dissipative qubit can induce heat exchange between the qubit and the measurement apparatus.}
    \label{fig:setup}
\end{figure}

In this paper, we focus on a qubit (two-level system), as the measured system and investigate the heat exchange between a qubit and a measurement apparatus when the qubit is coupled to bosonic environments, i.e., a dissipative qubit, under continuous measurement (see Fig.~\ref{fig:setup}).
By performing continuous measurements on an arbitrary pure state of the qubit, we identify the direction of steady-state heat flow and find lower and upper bounds on its magnitude.
Additionally, by shedding light on the transient dynamics of heat flow, we elucidate the excess heat transported by continuous quantum measurement.
We set $\hbar=1$ throughout this paper.

\section{model}

\subsection{Dissipative qubit}

The Hamiltonian that describes the qubit and $N$ bosonic heat baths is 
\begin{align}
H=H_{\rm qb}+\sum_{rk}\frac{\lambda_{rk}}{2}\left(\sigma_+b_{rk}+\sigma_-b^\dagger_{rk}\right)+\sum_{rk}\omega_{rk}b^\dagger_{rk}b_{rk}.
\end{align}
The qubit is represented by $H_{\rm qb}=(\Delta/2)\sigma_z$, where $\Delta$ is the energy difference between two levels and $\sigma_{x,y,z}$ are Pauli matrices.
When the qubit is weakly coupled to heat baths, the interaction between the qubit and $k$th mode in heat bath $r$ ($=1,2,\dots,N$) can be described by the Jaynes-Cummings-type coupling with coupling strength $\lambda_{rk}$.
Here, $\sigma_\pm=(\sigma_x\pm i\sigma_y)/2$ is the ``spin'' raising (lowering) operator of the qubit, and $b_{rk}$ and $b_{rk}^\dagger$ are the bosonic annihilation and creation operators of the $k$th mode in heat bath $r$, respectively.
The dissipation to heat bath $r$ via $\lambda_{rk}$ is characterized by the Ohmic spectral density~\cite{Leggett1987,Weiss_text}
\begin{align}
\label{eq:spectral}
I_r(\omega)
\equiv\sum_{k}\lambda_{rk}^2\delta(\omega-\omega_{rk})
=2\kappa_r\omega e^{-\omega/\omega_{\rm c}},
\end{align}
where $\kappa_r$ represents a dimensionless coupling strength to dissipative bath $r$ and $\omega_{\rm c}$ is the cutoff energy.
The bosonic baths are modeled as a collection of harmonic oscillators with the $k$th mode's energy $\omega_{rk}$.

\subsection{Continuous quantum measurement}

We consider continuous quantum measurement of an arbitrary pure state of the qubit,
\begin{align}
\ket{n}=\frac{c_g\ket{g}+c_e\ket{e}}{\sqrt{|c_g|^2+|c_e|^2}}
\end{align}
where $\ket{g}$ and $\ket{e}$ are the ground and excited states of the isolated qubit, respectively, and $c_{g,e}$ are complex coefficients.
The projection operator onto the pure state is expressed in terms of the identity operator $I$ and the Pauli operators $\{\sigma_z,\sigma_+,\sigma_-\}$ as 
\begin{align}
P_n=\ket{n}\bra{n}=I/2+\alpha\sigma_z+\beta\sigma_++\beta^*\sigma_-,
\end{align}
where the coefficients are
\begin{align}
\alpha=\frac{1}{2}\frac{|c_e|^2-|c_g|^2}{|c_e|^2+|c_g|^2}, \quad \beta=\frac{c_ec_g^*}{|c_e|^2+|c_g|^2},
\end{align}
and thus $\alpha^2+|\beta|^2=1/4$ holds.
The quantum dynamics of the density matrix $\varrho(t)$ of the global system (the qubit and the baths) is described by the quantum master equation~\cite{Carmichael_text,Wiseman2009_text,Bhandari2022},
\begin{align}
\dot{\varrho}(t)=-i[H,\varrho(t)]+\mathcal{D}_{\rm M}[\varrho(t)].
\end{align}
Here, the measurement effect is incorporated as
\begin{align}
\label{eq:DM}
\mathcal{D}_{\rm M}[\varrho(t)]=\gamma\left(P_n\varrho(t)P_n-\frac{1}{2}\left\{P_n,\varrho(t)\right\}\right),
\end{align}
where $\gamma$ is the measurement strength.
Note that the density matrix $\varrho(t)$ is ensemble averaged over the measurement outcomes.

\subsection{Lindblad equation}

When the qubit is weakly coupled to the heat baths ($\kappa_r\ll1$), by tracing out the degrees of freedom of the heat baths in local thermal equilibria with temperature $T_r$, we assume the Markov approximation and then obtain the Lindblad equation for the reduced density matrix $\rho(t)={\rm tr}_{\rm B}[\varrho(t)]$ as~\cite{Breuer_text,Ivander2022,Kamimura2023}
\begin{align}
\label{eq:lindblad}
\dot{\rho}(t)=-i[H_{\rm qb},\rho(t)]+\mathcal{D}_{\rm B}[\rho(t)]+\mathcal{D}_{\rm M}[\rho(t)],
\end{align}
where the dissipator is
\begin{align}
\mathcal{D}_{\rm B}[\rho]
&=\sum_r\Gamma^{\rm a}_r(\sigma_+\rho\sigma_--\frac{1}{2}\left\{\sigma_-\sigma_+,\rho\right\}) \nonumber\\
&+\sum_r\Gamma^{\rm e}_r(\sigma_-\rho\sigma_+-\frac{1}{2}\left\{\sigma_+\sigma_-,\rho\right\}).
\end{align}
The first and second terms of the dissipator $\mathcal{D}_{\rm B}$ represent single-photon absorption and emission processes with rates, \begin{align}
\Gamma^{\rm a}_r&=\frac{\pi}{2}I_r(\Delta)n_r(\Delta), \quad
\Gamma^{\rm e}_r&=\frac{\pi}{2}I_r(\Delta)[1+n_r(\Delta)], 
\end{align}
respectively, where $n_r(\varepsilon)=1/[e^{\varepsilon/(k_{\rm B}T_r)}-1]$ is the Bose-Einstein distribution function of heat bath $r$.
Note that since the quantum measurement~\eqref{eq:DM} acts only on the qubit Hilbert space, the contributions from the heat baths and the quantum measurement apparatus to the Lindblad equation are additive.

\subsection{Heat current}

The continuous quantum measurement changes the qubit state, which leads to heat exchange between the qubit and the measurement apparatus.
The heat current flowing out of the measurement apparatus into the qubit is given by~\cite{Ivander2022,Bhandari2022}
\begin{align}
J_{\rm M}(t)={\rm tr}_{\rm qb}\left[H_{\rm qb}\mathcal{D}_{\rm M}[\rho(t)]\right].
\end{align}
Note that from the definition of heat current, a positive $J_{\rm M}$ indicates that the qubit absorbs heat from the measurement apparatus, and a negative $J_{\rm M}$ indicates the opposite.
The heat current can be rewritten as
\begin{align}
\label{eq:heat_current}
J_{\rm M}(t)
=-\gamma\Delta|\beta|^2\braket{\sigma_z}_t
+\alpha\gamma\Delta\left(\beta\braket{\sigma_+}_t+\beta^*\braket{\sigma_-}_t\right),
\end{align}
where $\braket{\mathcal{O}}_t={\rm tr}_{\rm qb}[\rho(t)\mathcal{O}]$.

\section{steady-state heat current}

Let us consider the steady-state heat current, $J_{\rm M}=J_{\rm M}(t\to\infty)$, flowing out of the measurement apparatus into the qubit.
In the steady-state limit, we can solve the Lindblad equation~\eqref{eq:lindblad} and obtain the steady-state heat current as (see Appendix~\ref{app:1} for the details of the derivations)
\begin{align}
\label{eq:JM}
J_{\rm M}=\frac{|\beta|^2\Delta\gamma\Gamma_-\left[4\Delta^2+\Gamma_+(\Gamma_++\gamma)\right]}{4\Delta^2(\Gamma_++2|\beta|^2\gamma)+\Gamma_+(\Gamma_++\gamma)(\Gamma_++\gamma-2|\beta|^2\gamma)},
\end{align}
where $\Gamma_-=\sum_r\left(\Gamma^{\rm e}_r-\Gamma^{\rm a}_{r}\right)$ and $\Gamma_+
=\sum_r\left(\Gamma^{\rm e}_r+\Gamma^{\rm a}_{r}\right)$.
Since $0\le|\beta|^2\le1/4$, the steady-state heat current is bounded by
\begin{align}
\label{eq:JMrange}
0\le J_{\rm M} \le \frac{\Delta\gamma\Gamma_-}{4\Gamma_++2\gamma}=J_{\rm M}^{\rm max}.
\end{align}
This indicates that the qubit always absorbs heat from the measurement apparatus regardless of the measured state.

\begin{figure}[tb]
    \centering
    \includegraphics[width=1.0\columnwidth]{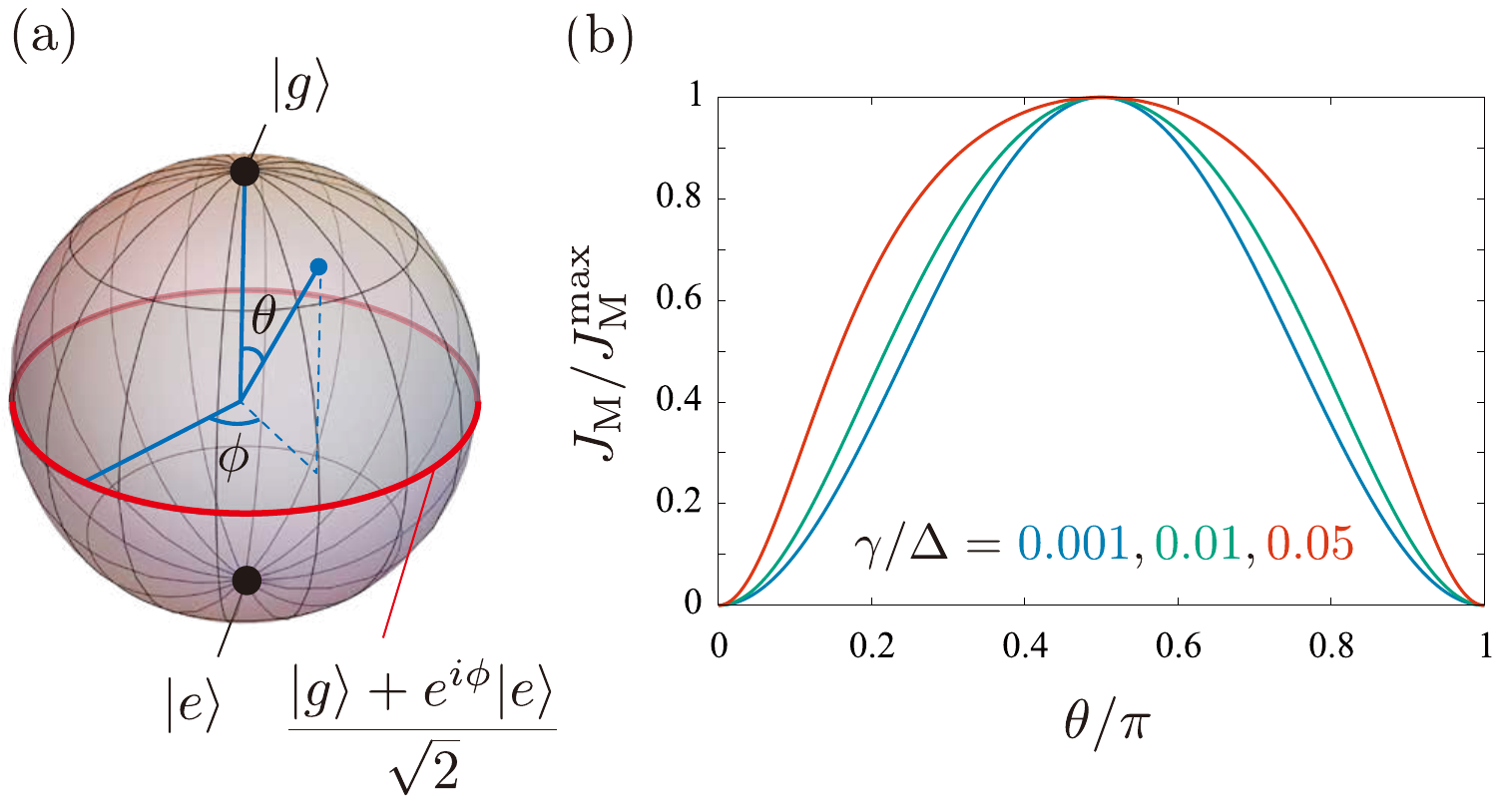}
    \caption{(a) Bloch sphere. (b) Steady-state heat current as a function of $\theta$ for $\Gamma_+/\Delta=0.01$ and $\gamma/\Delta=0.001$, 0.01, 0.05.}
    \label{fig:JMphi}
\end{figure}

The steady-state heat current~\eqref{eq:JM} depends only on $|\beta|^2$, i.e., $|c_g|$ and $|c_e|$, not on the relative phase difference between $c_g$ and $c_e$.
In the Bloch sphere representation (see Fig.~\ref{fig:JMphi}~(a)), the heat current depends on the latitude.
In Fig.~\ref{fig:JMphi}~(b), we show the dependence of the steady-state heat current on the zenith angle $\theta$.
The steady-state heat current vanishes at the north pole ($\theta=0$) and the south pole ($\theta=\pi$) and reaches a maximum at the equator ($\theta=\pi/2$).

\subsection{Lower bound ($\theta=0$, $\pi$)}

Now, let us focus on the lower and upper bounds of the steady-state heat current.
When the measurement apparatus monitors an eigenstate, $\ket{n}=\ket{g}$ or $\ket{e}$, ($\beta=0$), the heat current goes to 0, which is the lower bound.
In this case, the measurement effect~\eqref{eq:DM} reads
\begin{align}
\mathcal{D}_{\rm M}[\rho]=\frac{\gamma}{4}(\sigma_z\rho\sigma_z-\rho),
\end{align}
which is pure dephasing~\cite{Iyoda2013,Yamamoto2022}.
Since the pure dephasing acts only on the off-diagonal elements of the reduced density matrix on the eigenenergy basis, it does not contribute to the heat current.
Moreover, we can verify that no heat flows from the measurement apparatus not only in the case of a single qubit but also in the case of a general quantum system by monitoring an eigenstate of the isolated quantum system.

\begin{figure}[tb]
    \centering
    \includegraphics[width=1.0\columnwidth]{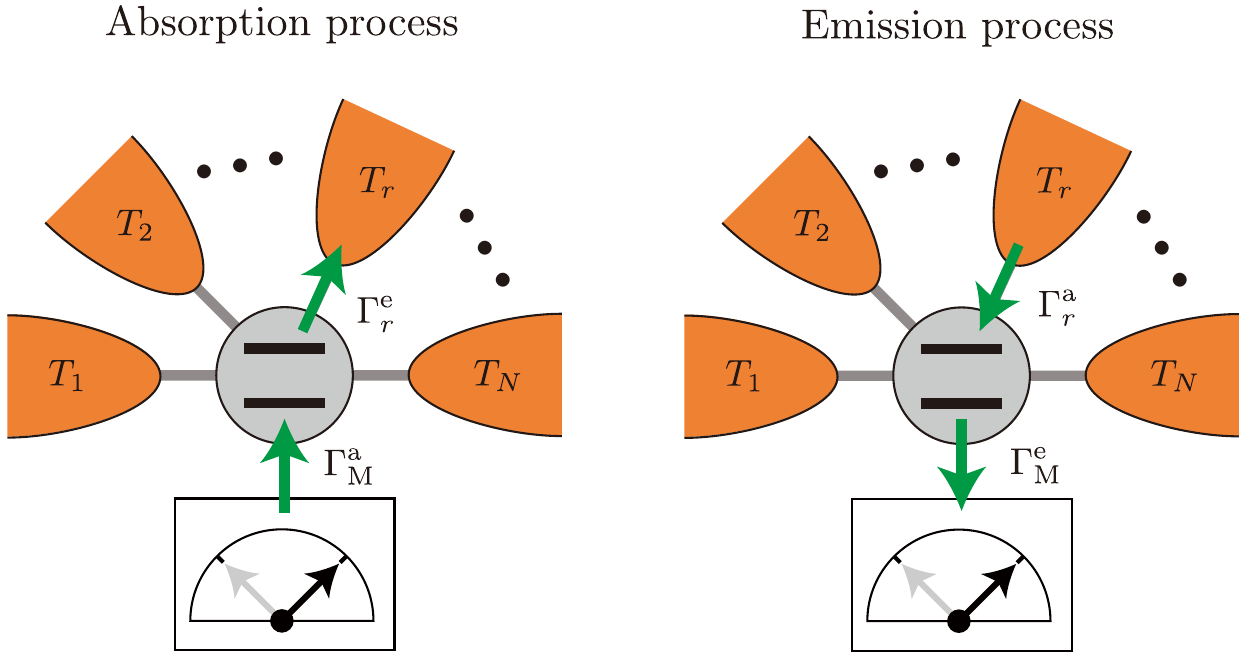}
    \caption{Elastic thermal photon processes. In the absorption process (left), a thermal photon with energy $\Delta$ travels from the measurement apparatus to the qubit and then to one of the heat baths. In the emission process (right), the opposite occurs.}
    \label{fig:JMmax}
\end{figure}

\subsection{Upper bound ($\theta=\pi/2$)}

The steady-state heat current reaches the upper bound $J_{\rm M}^{\rm max}$ when the measured state is a superposition state of the ground and excited states with the same probabilities, i.e., $\ket{n}=(\ket{g}+e^{i\phi}\ket{e})/\sqrt{2}$ ($|\beta|=1/2$).
In this measurement, $\mathcal{D}_{\rm M}[\rho]$ is composed of two parts, $\mathcal{D}_{\rm M}[\rho]=\mathcal{D}^{(1)}_{\rm M}[\rho]+\mathcal{D}^{(2)}_{\rm M}[\rho]$,
where
\begin{subequations}
\begin{align}
\mathcal{D}^{(1)}_{\rm M}[\rho]
&=\frac{\gamma}{4}\left(-\rho+\sigma_+\rho\sigma_-+\sigma_-\rho\sigma_+\right), \\
\mathcal{D}^{(2)}_{\rm M}[\rho]
&=\frac{\gamma}{4}\left(e^{-2i\phi}\sigma_+\rho\sigma_++e^{2i\phi}\sigma_-\rho\sigma_-\right).
\end{align}
\end{subequations}
Since the diagonal elements of $\mathcal{D}_{\rm M}^{(2)}[\rho(t)]$ are zero, it does not contribute to the heat current.
The remaining contribution $\mathcal{D}_{\rm M}^{(1)}[\rho]$ is of the same form as the dissipator $\mathcal{D}_{\rm B}[\rho]$ due to the heat baths with the absorption and emission rates,
\begin{align}
\Gamma_{\rm M}^{\rm a}=\frac{\gamma}{4}, \quad 
\Gamma_{\rm M}^{\rm e}=\frac{\gamma}{4},
\end{align}
respectively.
The same absorption and emission rates ($\Gamma_{\rm M}^{\rm a}=\Gamma_{\rm M}^{\rm e}$) indicate a heat bath with infinite temperature.
Note that for $\theta\ne\pi/2$ the measurement effect to the quantum dynamics~\eqref{eq:DM} is not the same form of the dissipator of the infinite-temperature bath.
We can understand the upper bound of the steady-state heat current $J_{\rm M}^{\rm max}$ by considering elastic traveling of a thermal photon with the energy $\Delta$ (see Fig.~\ref{fig:JMmax}).
This picture is justified in that the coherence vanishes in the steady state limit, and then only the diagonal elements of the reduced density matrix on the eigenenergy basis remain.
The rate of the absorption (emission) process in which the thermal photon travels from (into) the measurement apparatus into (from) one of the heat baths via the qubit is given by $\gamma\sum_r\Gamma_{r}^{{\rm e}({\rm a})}/(4\Gamma_++2\gamma)$.
Therefore, the net heat current flowing out of the measurement apparatus reproduces the upper bound of the heat current.

Here, let us examine the case of the general quantum system under continuous quantum measurement onto a superposition state of two eigenstates, $\ket{n}=(\ket{i}+e^{i\phi}\ket{j})/\sqrt{2}$, where $\ket{i}$ and $\ket{j}$ are non-degenerate energy eigenstates of an isolated quantum system, $H_{\rm S}\ket{i}=E_i\ket{i}$.
The heat current flowing out of the measurement apparatus is given as
\begin{align}
J_{\rm M}(t)=\sum_{k}E_k\bra{k}\mathcal{D}_{\rm M}[\rho(t)]\ket{k}.
\end{align}
This expression indicates that only the diagonal elements of $\mathcal{D}_{\rm M}[\rho]$ contribute to the heat current, as we mentioned above.
We thus obtain
\begin{align}
\label{eq:JM_general}
J_{\rm M}(t)=-\frac{\gamma}{4}\left(E_i-E_j\right)\left[\rho_{ii}(t)-\rho_{jj}(t)\right],
\end{align}
where $\rho_{ij}(t)=\bra{i}\rho(t)\ket{j}$.
This expression suggests a nontrivial result.
The heat current is still positive unless the population inversion occurs between measured states~\cite{Higgins2014,Kloc2021,Ueki2022}.
However, in the case of the population inversion, such as the $\Lambda$ model~\cite{Hegerfeldt1993,Liu2001,Ou2008}, we can find a negative $J_{\rm M}$ ($<0$) even in the steady-state limit, which never happens in the single qubit case (see Appendix~\ref{app:2} for the details of the $\Lambda$ model as an example of the population inversion).
We stress that the interplay between the environment and quantum measurements results in the heat backflow.
The negative heat current indicates quantum measurement cools down the quantum system, whereas the qubit experiences heating up by measurement.
This cooling characteristic is useful for implementing a refrigerator.
For the qubit case, $E_{i(j)}=\mp\Delta/2$ and $\rho_{ii(jj)}=(1\mp\braket{\sigma_z})/2$.
Since $\braket{\sigma_z}=-\Gamma_-/(\Gamma_++\gamma/2)$ in the steady-state limit, the heat current is consistent with the upper bound $J_{\rm M}^{\rm max}$, and it is always positive.

\section{transient dynamics}

\begin{figure}[tb]
    \centering
    \includegraphics[width=1.0\columnwidth]{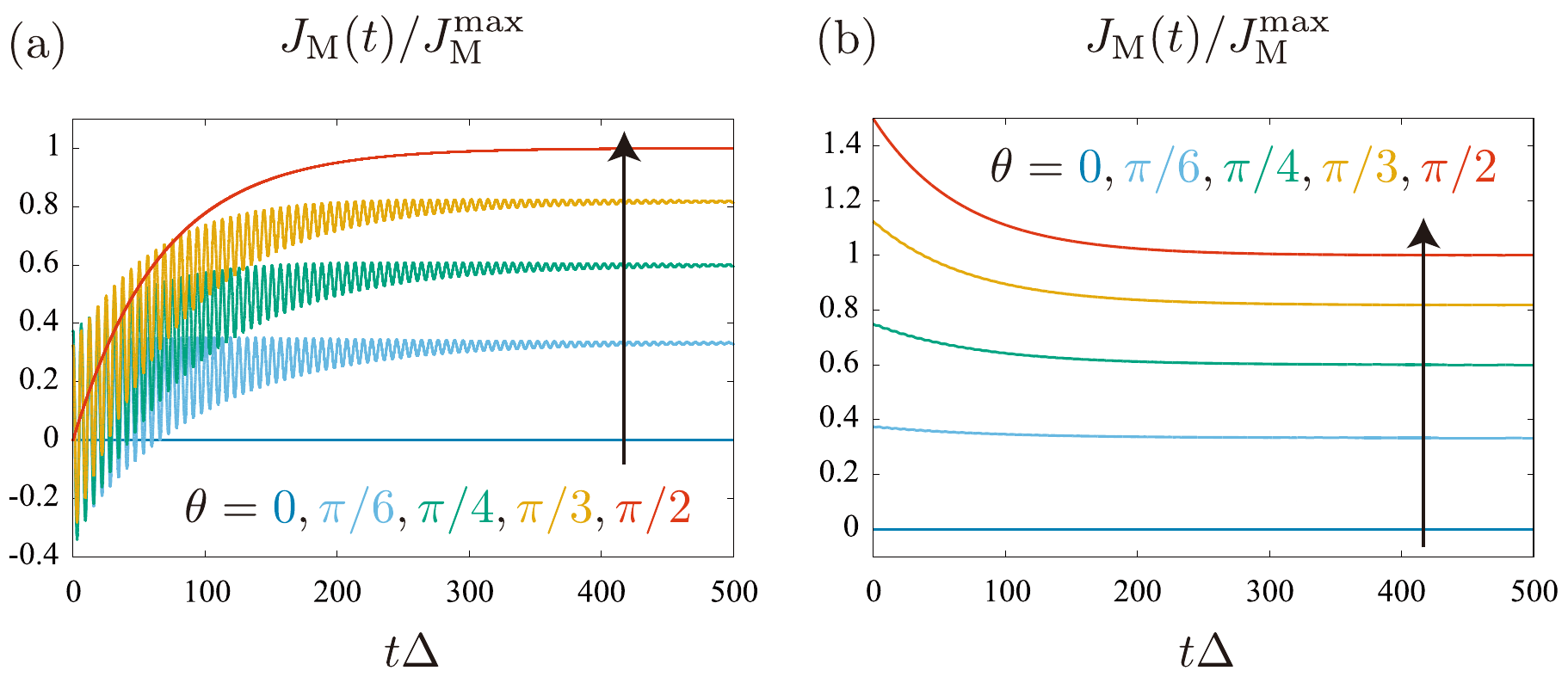}
    \caption{Dynamics of the heat current for $\Gamma_+=0.02\Delta$, $\Gamma_-=\gamma=0.01\Delta$, $\phi=0$, and $\theta=0$ (blue), $\pi/6$ (light blue), $\pi/4$ (green), $\pi/3$ (orange), $\pi/2$ (red). The initial state is prepared with  (a) $\braket{\sigma_x}_0=1$, $\braket{\sigma_y}_0=\braket{\sigma_z}_0=0$ and (b) $\braket{\sigma_x}_0=\braket{\sigma_x}_{\rm ss}=0$, $\braket{\sigma_y}_0=\braket{\sigma_y}_{\rm ss}=0$, $\braket{\sigma_z}_0=\braket{\sigma_z}_{\rm ss}=-\Gamma_-/\Gamma_+$.}
    \label{fig:JMtransient}
\end{figure}

In this section, we consider how the heat current in the transient regime reaches the steady state after switching on the measurement to a dissipative qubit.
The Lindblad equation~\eqref{eq:lindblad} can be analytically solved for two cases: $\beta=0$ and $|\beta|=1/2$, which provide lower and upper bounds for the steady-state heat current, respectively.
For $\beta=0$, the qubit does not exchange heat with the measurement apparatus, even in the transient regime, regardless of the preparation of the qubit.
For $|\beta|=1/2$, the heat current is given by (see Appendix~\ref{app:1} for the details of the derivation)
\begin{align}
J_{\rm M}(t)
=\frac{\gamma\Delta}{4}\left[\frac{\Gamma_-}{\tilde{\Gamma}_+}-\left(\braket{\sigma_z}_0+\frac{\Gamma_-}{\tilde{\Gamma}_+}\right)e^{-\tilde{\Gamma}_+t}\right],
\end{align}
where $\tilde{\Gamma}_+=\Gamma_++\gamma/2$.
The heat current exponentially changes at a rate $\tilde{\Gamma}_+$ and reaches the steady-state value.
Note that we observe the negative heat current in the transient regime due to the measurement effect on quantum coherence, which never happens in the steady-state limit, as shown in Fig.~\ref{fig:JMtransient}.
Particularly, for $|\beta|=1/2$, the transient heat current is larger than the maximum steady-state heat current when $\braket{\sigma_z}_0<-\Gamma_-/(\Gamma_++\gamma/2)$.

The heat current oscillates with time in the transient regime except for the analytically solvable points ($\theta=0$ and $\pi/2$) and then reaches the steady-state value, as shown in Fig.~\ref{fig:JMtransient}.
The oscillation comes from the coherence, $\braket{\sigma_x}_t$ and $\braket{\sigma_y}_t$, and its period is approximately $2\pi/\Delta$ because the qubit is weakly coupled to the heat baths and the measurement apparatus.

\begin{figure}[t]
    \centering
    \includegraphics[width=1.0\columnwidth]{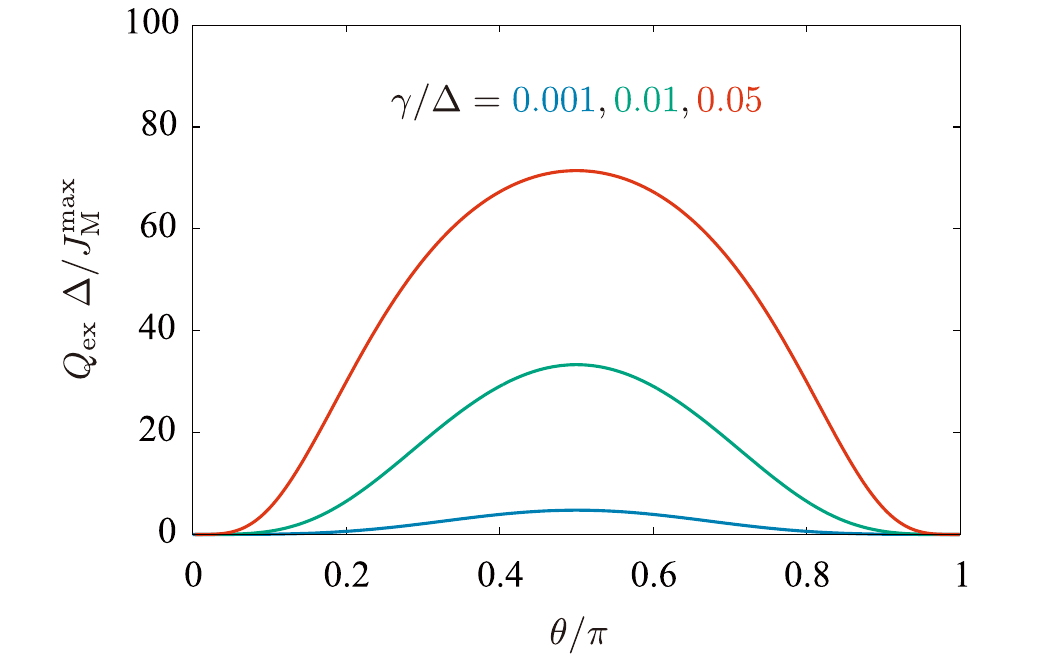}
    \caption{Excess heat flowing from the measurement apparatus into the qubit as a function of $\theta$ for $\Gamma_+=0.02\Delta$, $\Gamma_-=0.01\Delta$, and $\phi=0$.}
    \label{fig:Qex}
\end{figure}

When a continuous quantum measurement is made on the qubit in the steady state, the heat exchange between the measurement apparatus and the qubit begins, and then the qubit enters a new steady state in the presence of the quantum measurement, as shown in Fig.~\ref{fig:JMtransient}~(b).
The excess heat flowing from the measurement apparatus to the qubit during the transient process, starting from the steady state in the absence of the measurement, is given by
\begin{align}
Q_{\rm ex}=\int_0^\infty dt~\left[J_{\rm M}(t)-J_{\rm M}\right].
\end{align}
Figure~\ref{fig:Qex} shows the dependence of the excess heat on $\theta$.
Here, the excess heat is non-negative regardless of the measured state.
In the steady-state limit in the absence of a continuous quantum measurement, the coherence vanishes, and the population remains $\braket{\sigma_z}=-\Gamma_-/\Gamma_+$.
Therefore, just after switching on the measurement, the heat current is $J_{\rm M}(t=0)=\gamma\Delta|\beta|^2\Gamma_-/\Gamma_+$, and it is larger than $J_{\rm M}$ for any $\beta$ (the measured state), which results in $Q_{\rm ex}\ge0$.
The excess heat vanishes at $\theta=0$ and $\pi$ and reaches a maximum value at $\theta=\pi/2$.
The maximum excess heat at $\theta=\pi/2$ is
\begin{align}
Q_{\rm ex}^{\rm max}=\Delta\gamma\Gamma_{-}\frac{\Gamma_{+}^{-1}-\tilde{\Gamma}_{+}^{-1}}{4\tilde{\Gamma}_+}.
\end{align}
This can be understood as follows: Once the continuous quantum measurement is switched on, the energy stored in the qubit increases from $-\Delta\Gamma_-/(2\Gamma_+)$ to $-\Delta\Gamma_-/(2\tilde{\Gamma}_+)$.
The sources for supplying the energy to the qubit are the heat baths and the measurement apparatus, and the rate from the measurement apparatus is $\gamma/(2\tilde{\Gamma}_+)$.

\section{experimental feasibility}

We note that our theoretical setup is feasible using a platform of superconducting circuits.
The recent technological development allows experimentalists to detect the heat current through the superconducting elements, including a superconducting transmon qubit, between the heat baths, made of copper resistors, using the standard thermometric technique in the low-temperature range $T\gtrsim\hbar\Delta/k_{\rm B}\approx50~{\rm mK}$~\cite{Ronzani2018,Senior2020,Maillet2020}.
The advantage of our setup is the ability to access the heat current $J_{\rm M}$ by measuring the heat currents in each heat bath, thanks to energy conservation.
Moreover, continuous quantum measurement has already been experimentally demonstrated in a superconducting transmon qubit by performing a sequence of weak quantum measurements~\cite{Vijay2012,Murch2013,Weber2014}.

\section{summary}

We studied the heat flow between the measurement apparatus and a dissipative qubit under continuous measurement by using the Lindblad equation.
In the steady-state limit, we found that the heat current always flows out of the measurement apparatus into the qubit regardless of the measured qubit state; i.e., the measurement apparatus works as a heater of the qubit.
We identified lower and upper bounds for the steady-state heat current when the measured qubit state is respectively at the north or south poles ($\theta=0$ or $\pi$) and on the equator ($\theta=\pi/2 $) of the Bloch sphere and clarified the heat transfer process at the two bounds.
For the system with population inversion, we found that the heat current can flow in the opposite direction to the qubit even at the steady-state limit.
Furthermore, we considered the transient dynamics of the heat current, and by tracking the dynamics, we could calculate the excess heat induced by a continuous quantum measurement and found it to be positive; that is, the measurement apparatus supplies heat to the qubit.

This study helps our understanding of quantum thermodynamics in dissipative quantum systems by shedding light on the heat flow between the measurement apparatus and the measured quantum system.
Although we dealt with a qubit as the measured quantum system in this work, it remains a problem to identify the direction of heat flow in the general quantum case.

\section*{acknowledgement}

We thank S. Kamimura for fruitful discussion. The authors acknowledge support from the JST Moonshot R\&D–MILLENNIA program (Grant No. JPMJMS2061), and Y. T. acknowledges support from JSPS KAKENHI (Grant No. 23K03273).

\appendix

\section{Exact solutions of the Lindblad equation}
\label{app:1}

Here, we provide exact solutions of the Lindblad equation~\eqref{eq:lindblad} in the main text.

When expressing the density matrix in terms of the Bloch vectors as
\begin{align}
\rho(t)=\frac{1}{2}\left(I+\braket{\sigma_x}_t\sigma_x+\braket{\sigma_y}_t\sigma_y+\braket{\sigma_z}_t\sigma_z\right),
\end{align}
the Lindblad equation turns into three differential equations,
\begin{align}
\braket{\dot{\sigma}_z}_t
=&-\Gamma_--(\Gamma_++2|\beta|^2\gamma)\braket{\sigma_z}_t \nonumber\\
&+2\alpha\beta'\gamma\braket{\sigma_x}_t-2\alpha\beta''\gamma\braket{\sigma_y}_t, \\
\braket{\dot{\sigma}_x}_t
=&-\left(\frac{\Gamma_++\gamma}{2}-2\beta'^2\gamma\right)\braket{\sigma_x}_t \nonumber\\
&-\left(\Delta-2\beta'\beta''\gamma\right)\braket{\sigma_y}_t+2\alpha\beta'\gamma\braket{\sigma_z}_t, \\
\braket{\dot{\sigma}_y}_t
=&-\left(\frac{\Gamma_++\gamma}{2}+2\beta'^2\gamma\right)\braket{\sigma_y}_t \nonumber\\
&+\left(\Delta-2\beta'\beta''\gamma\right)\braket{\sigma_x}_t-2\alpha\beta''\gamma\braket{\sigma_z}_t.
\end{align}
where $\beta=\beta'+i\beta''$.
We note that each component decays at a rate equal to or greater than that in the absence of measurement ($\gamma=0$) because of $|\beta|^2\ge0$ and $0\le\beta'^2\le1/4$. This fact allows us to interpret continuous quantum measurement to a dissipative qubit as the anti-Zeno effect~\cite{Koshino2005}.

\subsection{A steady-state solution}

In the steady-state limit, $\braket{\dot{\sigma}_i}_{t\to\infty}=0$, we obtain
\begin{align}
\label{eq:zss}
\braket{\sigma_z}
=&\frac{-\Gamma_-[4\Delta^2+(\Gamma_++\gamma)(\Gamma_++\gamma-4|\beta|^2\gamma)]}{4\Delta^2(\Gamma_++2|\beta|^2\gamma)+\Gamma_+(\Gamma_++\gamma)(\Gamma_++\gamma-2|\beta|^2\gamma)},\\
\label{eq:xss}
\braket{\sigma_x}
=&\frac{-4\alpha\gamma\Gamma_-[2\Delta\beta''+(\Gamma_++\gamma)\beta']}{4\Delta^2(\Gamma_++2|\beta|^2\gamma)+\Gamma_+(\Gamma_++\gamma)(\Gamma_++\gamma-2|\beta|^2\gamma)},\\
\label{eq:yss}
\braket{\sigma_y}
=&\frac{-4\alpha\gamma\Gamma_-[2\Delta\beta'+(\Gamma_++\gamma)\beta'']}{4\Delta^2(\Gamma_++2|\beta|^2\gamma)+\Gamma_+(\Gamma_++\gamma)(\Gamma_++\gamma-2|\beta|^2\gamma)}.
\end{align}
Note that the coherences remain due to the continuous quantum measurement, even in the steady-state limit.
From the expression of the heat current~\eqref{eq:heat_current}, the steady-state heat current can be rewritten as
\begin{align}
J_{\rm M}
=&-\gamma\Delta|\beta|^2\braket{\sigma_z}+\alpha\gamma\Delta(\beta'\braket{\sigma_x}-\beta''\braket{\sigma_y}),
\end{align} 
and, by substituting the steady-state solution~\eqref{eq:zss}-\eqref{eq:yss}, we obtain Eq.~\eqref{eq:JM} in the main text.

\subsection{Transient dynamics}

In general, the three differential equations in which all components are mixed cannot be analytically solved.
However, since two of the three differential equations are closed, we can exactly solve the Lindblad equation for two cases: (i) $\alpha=1/2$, $\beta=0$ and (ii) $\alpha=0$, $|\beta|=1/2$.

\subsubsection*{(i) Case of $\alpha=1/2$, $\beta=0$}

This case corresponds to a continuous quantum measurement on an eigenstate of the qubit.
The Lindblad equation reduces to
\begin{align}
\braket{\dot{\sigma}_z}_t
=&-\Gamma_--\Gamma_+\braket{\sigma_z}_t, \\
\braket{\dot{\sigma}_x}_t
=&-\frac{\Gamma_++\gamma}{2}\braket{\sigma_x}_t-\Delta\braket{\sigma_y}_t, \\
\braket{\dot{\sigma}_y}_t
=&-\frac{\Gamma_++\gamma}{2}\braket{\sigma_y}_t+\Delta\braket{\sigma_x}_t.
\end{align}
We can solve it easily:
\begin{align}
\braket{\sigma_z}_t
=&-\frac{\Gamma_-}{\Gamma_+}+\left(\braket{\sigma_z}_0+\frac{\Gamma_-}{\Gamma+}\right)e^{-\Gamma_+t}, \\
\braket{\sigma_x}_t
=&e^{-(\Gamma_++\gamma)t/2}\left[\braket{\sigma_x}_0\cos(\Delta t)-\braket{\sigma_y}_0\sin(\Delta t)\right], \\
\braket{\sigma_y}_t
=&e^{-(\Gamma_++\gamma)t/2}\left[\braket{\sigma_x}_0\sin(\Delta t)+\braket{\sigma_y}_0\cos(\Delta t)\right].
\end{align}

\subsubsection*{(ii) Case of $\alpha=0$, $|\beta|=1/2$}

In this case, the measurement apparatus monitors the superposition state of the qubit.
Without loss of generality, $\beta$ can be written as $\beta=e^{i\phi}/2$.
The Lindblad equation reads
\begin{align}
\braket{\dot{\sigma}_z}_t
=&-\Gamma_--\left(\Gamma_++\frac{\gamma}{2}\right)\braket{\sigma_z}_t, \\
\braket{\dot{\sigma}_x}_t
=&-\frac{2\Gamma_++\gamma-\gamma\cos2\phi}{4}\braket{\sigma_x}_t \nonumber\\
&-\left(\Delta-\frac{\gamma}{4}\sin 2\phi\right)\braket{\sigma_y}_t, \\
\braket{\dot{\sigma}_y}_t
=&-\frac{2\Gamma_++3\gamma+\gamma\cos2\phi}{4}\braket{\sigma_y}_t \nonumber\\
&+\left(\Delta-\frac{\gamma}{4}\sin 2\phi\right)\braket{\sigma_x}_t,
\end{align}
and its solution is given as
\begin{align}
\braket{\sigma_z}_t
=&-\frac{\Gamma_-}{\Gamma_++\gamma/2}+\left(\braket{\sigma_z}_0+\frac{\Gamma_-}{\Gamma_++\gamma/2}\right)e^{-(\Gamma_++\gamma/2)t}, \\
\braket{\sigma_x}_t
=&\frac{e^{-(\Gamma_++\gamma)t/2}}{2\delta}\Big[(\delta\cos\delta t+\gamma\cos^2\phi \sin\delta t)\braket{\sigma_x}_0 \nonumber\\
&\qquad\qquad\quad -2\left(\Delta-\frac{\gamma}{4}\sin2\phi\right)\sin\delta t\braket{\sigma_y}_0\Big], \\
\braket{\sigma_y}_t
=&\frac{e^{-(\Gamma_++\gamma)t/2}}{2\delta}\Big[(\delta\cos\delta t-\gamma\cos^2\phi\sin\delta t)\braket{\sigma_y}_0 \nonumber\\
&\qquad\qquad\quad -2\left(\Delta-\frac{\gamma}{4}\sin2\phi\right)\sin\delta t \braket{\sigma_x}_0\Big],
\end{align}
where $\delta=\sqrt{[\Delta-(\gamma/4)\sin2\phi]^2-(\gamma/2)^2\cos^4\phi}$ is real because of $\Delta\gg\gamma$.

\section{Example of population inversion: the $\Lambda$ model}
\label{app:2}

\begin{figure}[tb]
    \centering
    \includegraphics[width=1.0\columnwidth]{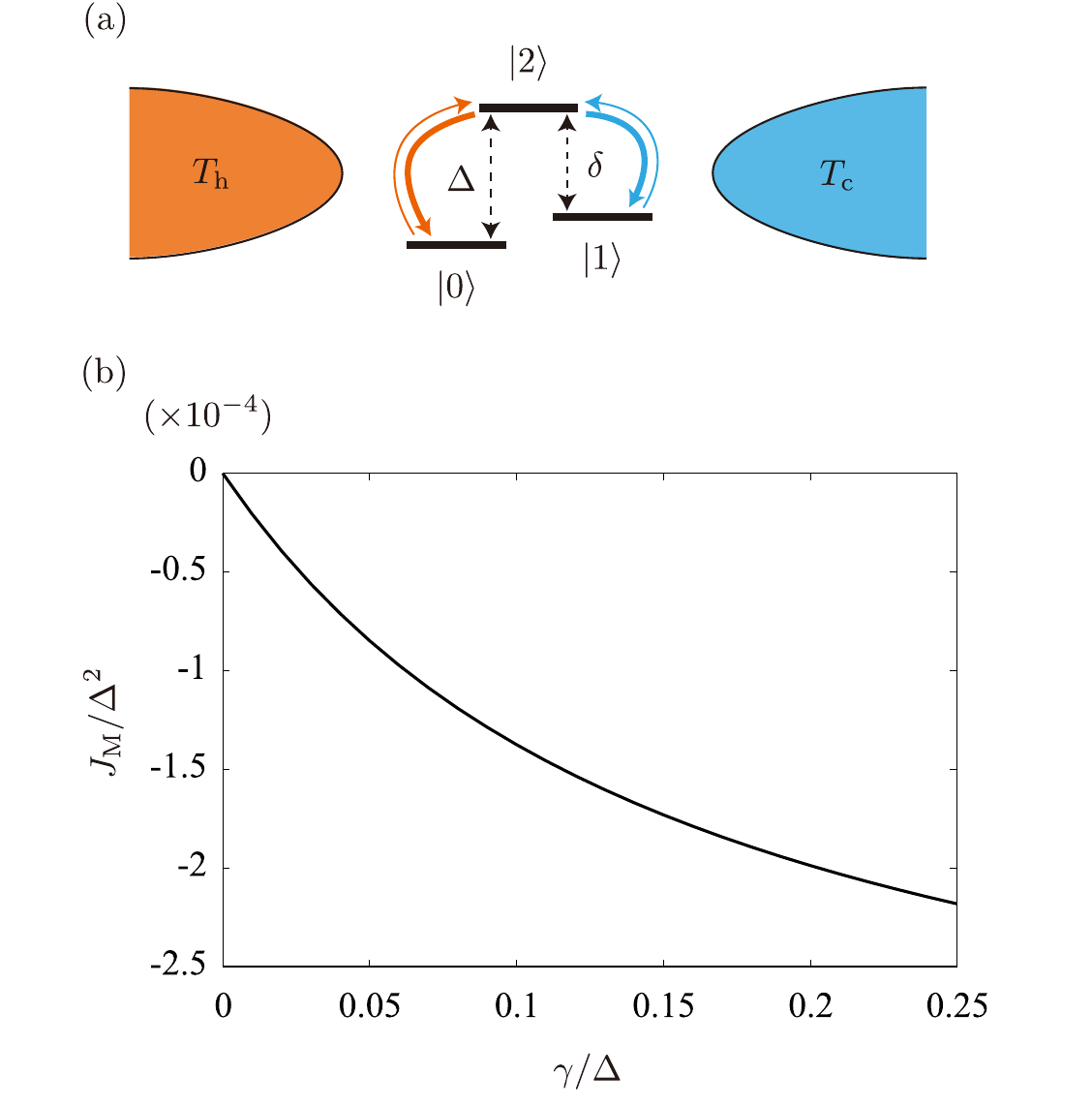}
    \caption{(a) Schematic diagram of the $\Lambda$ model with energy differences $\Delta$ and $\delta$ ($<\Delta$).
    The three-level system is coupled to the two bosonic heat baths ($T_{\rm h}>T_{\rm c}$).
    The hot bath allows transitions between $\ket{0}$ and $\ket{2}$, while the cold bath allows them between $\ket{1}$ and $\ket{2}$.
    (b) The steady-state heat current $J_{\rm M}$ as a function of $\gamma$ under the measurement on $(\ket{0}+\ket{1})/\sqrt{2}$ for $\delta/\Delta=0.5$, $k_{\rm B}T_{\rm h}/\Delta=5$, $k_{\rm B}T_{\rm c}/\Delta=2$, and $\kappa_{\rm h}=\kappa_{\rm c}=0.01$.}
    \label{fig:lambda}
\end{figure}

Here, we show that the steady-state heat current can be negative for the $\Lambda$ model in which a population inversion can occur.
The $\Lambda$ model is a three-level system coupled to two bosonic heat baths, as shown in Fig.~\ref{fig:lambda}~(a).
Its Hamiltonian is given by
\begin{align}
H=H_{\rm S}+\sum_{r={\rm h},{\rm c}}H_{{\rm B},r}+\sum_{r={\rm h},{\rm c}}H_{{\rm int},r}.
\end{align}
The three-level system is described by 
\begin{align}
H_{\rm S}=\sum_{i=0,1,2}\varepsilon_i\ket{i}\bra{i},
\end{align}
where $\varepsilon_i$ are eigenenergies of the isolated three-level system, $\varepsilon_0=0$, $\varepsilon_1=\Delta-\delta$, and $\varepsilon_2=\Delta$ with $\varepsilon_0<\varepsilon_1<\varepsilon_2$.
The hot and cold bath $H_{{\rm B},{\rm h}({\rm c})}$ are bosonic heat baths, of the same form as in the qubit case in the main text with $T_{\rm h}>T_{\rm c}$.
The hot bath is coupled to $\ket{0}$ and $\ket{2}$ of the three-level system, while the cold bath is coupled to $\ket{1}$ and $\ket{2}$,
\begin{align}
H_{{\rm int},{\rm h}}
&=\sum_k\frac{\lambda_{{\rm h},k}}{2}\left(\ket{2}\bra{0}b_{{\rm h},k}+\ket{0}\bra{2}b^\dagger_{{\rm h},k}\right), \\
H_{{\rm int},{\rm c}}
&=\sum_k\frac{\lambda_{{\rm c},k}}{2}\left(\ket{2}\bra{1}b_{{\rm c},k}+\ket{1}\bra{2}b^\dagger_{{\rm c},k}\right),
\end{align}
where $\lambda_{r,k}$ is the coupling strength.
Similar to the qubit case in the main text, we introduce the Ohmic spectral density~\eqref{eq:spectral}.

The quantum dynamics of the reduced density matrix $\rho(t)$ can be described by the Lindblad equation,
\begin{align}
\label{eq:lindblad_lambda}
\dot{\rho}(t)
=-i[H_{\rm S},\rho(t)]+\mathcal{D}_{\rm B}[\rho(t)]+\mathcal{D}_{\rm M}[\rho(t)],
\end{align}
where the continuous quantum measurement contribution $\mathcal{D}_{\rm M}[\rho(t)]$ takes the same form as Eq.~(4).
Here, the dissipation due to the heat baths is represented by
\begin{align}
\mathcal{D}_{\rm B}[\rho(t)]
&=\Gamma_{\rm h}^{\rm e}\left[\ket{0}\bra{2}\rho(t)\ket{2}\bra{0}-\frac{1}{2}\left\{\ket{2}\bra{2},\rho(t)\right\}\right] \nonumber\\
&+\Gamma_{\rm h}^{\rm a}\left[\ket{2}\bra{0}\rho(t)\ket{0}\bra{2}-\frac{1}{2}\left\{\ket{0}\bra{0},\rho(t)\right\}\right] \nonumber\\
&+\Gamma_{\rm c}^{\rm e}\left[\ket{1}\bra{2}\rho(t)\ket{2}\bra{1}-\frac{1}{2}\left\{\ket{2}\bra{2},\rho(t)\right\}\right] \nonumber\\
&+\Gamma_{\rm c}^{\rm a}\left[\ket{2}\bra{1}\rho(t)\ket{1}\bra{2}-\frac{1}{2}\left\{\ket{1}\bra{1},\rho(t)\right\}\right],
\end{align}
where the emission and absorption rates of the hot and cold baths are
\begin{align}
\Gamma_{\rm h}^{\rm e}&=\frac{\pi}{2}I_{\rm h}(\Delta)[1+n_{\rm h}(\Delta)], \quad
\Gamma_{\rm h}^{\rm a}=\frac{\pi}{2}I_{\rm h}(\Delta)n_{\rm h}(\Delta), \nonumber\\
\Gamma_{\rm c}^{\rm e}&=\frac{\pi}{2}I_{\rm c}(\delta)[1+n_{\rm c}(\delta)], \quad~~~
\Gamma_{\rm c}^{\rm a}=\frac{\pi}{2}I_{\rm c}(\delta)n_{\rm c}(\delta),
\end{align}
respectively.
The rate imbalance induces a population inversion between $\ket{0}$ and $\ket{1}$ when $\Delta/T_{\rm h}<\delta/T_{\rm c}$ in the absence of a continuous quantum measurement~\cite{Ueki2022}.
If the population inversion still occurs when performing the continuous quantum measurement on the superposition state of $\ket{0}$ and $\ket{1}$, i.e., $(\ket{0}+e^{i\phi}\ket{1})/\sqrt{2}$, the general expression of the heat current~\eqref{eq:JM_general} suggests $J_{\rm M}<0$.

Figure~\ref{fig:lambda}~(b) shows the $\gamma$ dependence of the steady-state heat current flowing from the measurement apparatus into the three-level system under continuous quantum measurement of $(\ket{0}+\ket{1})/\sqrt{2}$.
We interestingly find that the heat current is negative in the steady-state limit regardless of $\gamma$, as a result of the population inversion between the measured states.
Note that we find that the steady-state heat current is independent of $\phi$ similarly to the qubit case.

\bibliography{heat_measurement_apparatus}

\end{document}